# Physics-informed Reduced-Order Learning from the First Principles for Simulation of Quantum Nanostructures

Martin Veresko and Ming-Cheng Cheng*

Department of Electrical and Computer Engineering

Clarkson University, Potsdam, NY 13699-5720, USA

vereskm@clarkson.edu;  *mcheng@clarkson.edu

## Abstract

Multi-dimensional direct numerical simulation (DNS) of the Schrödinger equation is needed for design and analysis of quantum nanostructures that offer numerous applications in biology, medicine, materials, electronic/photonic devices, etc. In large-scale nanostructures, extensive computational effort needed in DNS may become prohibitive due to the high degrees of freedom (DoF). This study employs a physics-based reduced-order learning algorithm, enabled by the first principles, for simulation of the Schrödinger equation to achieve high accuracy and efficiency. The proposed simulation methodology is applied to investigate two quantum-dot structures; one operates under external electric field, and the other is influenced by internal potential variation with periodic boundary conditions. The former is similar to typical operations of nanoelectronic devices, and the latter is of interest to simulation and design of nanostructures and materials, such as applications of density functional theory. In each structure, cases within and beyond training conditions are examined. Using the proposed methodology, a very accurate prediction can be realized with a reduction in the DoF by more than 3 orders of magnitude and in the computational time by 2 orders, compared to DNS. An accurate prediction beyond the training conditions, including higher external field and larger internal potential in untrained quantum states, is also achieved. Comparison is also carried out between the physics-based learning and Fourier-based plane-wave approaches for a periodic case.

**Introduction**

In recent years, reduced order modeling techniques have been widely adopted to decrease the computational complexity in scientific and engineering problems [1-12]. Many of these techniques involve projecting the problem of interest onto a set of orthogonal basis functions to reduce the degrees of freedom (DoF) using the basis functions that are optimal for the problem. Proper orthogonal decomposition (POD) [13,14] is one of the such modeling techniques, where its basis functions (hereafter referred to as POD modes) are generated by maximizing the mean square inner product of each mode with the solution data for the problem [2-7]. The solution data used in the decomposition should incorporate as much as possible a realistic range of parametric variations for the problem. This data-driven *learning* process generates an optimal set of POD modes for the problem of interest to minimize the DoF with a smallest possible least square (LS) error if the quality of the collected data is adequate. This is similar to principal component analysis (PCA) [11,12] and single value decomposition (SVD) [8-10] but different from many other projection-based techniques that prescribe a set of assumed basis functions based on the solution form of the problem imposed by the geometry, excitations and/or boundary conditions (BCs) of the problem. These projection methods include, for example, Fourier basis, Legendre polynomials, Bessel functions, Airy functions, Wannier functions, etc., each of which only works well for specific scientific or engineering problems whose solution forms are well-defined and close to the selected set of basis functions.

The above-mentioned maximization process trains an optimal set of POD modes to acquire the essential information embedded in the solution data. To derive a simulation approach that complies with the physical principles of the problem, Galerkin projection of the governing equation for the problem onto a mathematical space constituted by these modes is applied. This is a crucial step to incorporate clear physical principles in this POD-Galerkin simulation methodology in order to reach high accuracy with only a handful of DoF. This is very different from most machine learning methods solving physical quantities without the governing equations of the problems [15-20], which rely only on statistical learning approaches to minimize the deviation of the prediction without accounting for physical principles.

The concept of POD has been applied to many research areas including fluid dynamics, heat conduction, power distributions and some other problems governed by partial differential equations (PDEs) [3-6,17-23]. Many of these studies implemented machine learning, such as neural networks [17,18] or radial basis functions [19,20], without incorporating the governing equation of the system. In this work, the POD-Galerkin methodology is applied to investigate its capability of predicting accurate wave functions (WFs) in nanostructures with high efficiency by solving the Schrödinger equation. Electronic/photonic nanostructures and nanoscale devices are the backbone of the modern photonic, electronic and computing technologies including lasers [24,25], light emitting diodes (LEDs) [26-29], photovoltaics [30,31], nanoscale transistors [32-35], etc. To understand and design these nanoscale structures/devices, electron/hole WFs and eigenenergies need to be solved from the Schrödinger equation subjected to an internal/external electric field or potential. Another important application of the quantum eigenvalue problems is electronic structure calculations using density functional theory (DFT) in many areas of research and industries, such as, pharmaceutical medicine, electronics, photonics, food science, material science, etc. [36-45]. More recently, non-equilibrium Green's functions [46] have been applied to develop quantum transport models to improve the accuracy of multi-dimensional nanoscale device simulations that have traditionally been based on semiclassical transport models. To account for atomic-scale effects in quantum transport models, *ab initio* DFT simulations are needed to provide the electronic structure and Hamiltonian and to evaluate material and transport parameters for the nanoscale devices [47-49]. DFT simulations however demand immense computational resources for solving the Kohn-Sham equations, a set of Schrödinger-like equations for many electrons in the structure of interest in which a modified effective potential for each electron encapsulates the interactions of the many-body system. Efficiency and accuracy of multi-dimensional simulations of quantum eigenvalue problems are thus crucial for further advance in these emerging nanoscale structures/devices.

To offer an effective approach to solve the Schrödinger equation, two physics-informed POD-Galerkin simulation techniques, individual [7,50] and global [50] approaches, were developed and investigated for simulations of the Schrödinger equation in 1D [7] and 2D [50] nanostructures. The individual approach trains the POD modes using the WF solution data collected in each individual quantum state (QS) from direct numerical simulations (DNSs) of the Schrödinger equation for the structure of interest [7,50]. There are therefore $N_Q$ sets of trained POD modes needed for $N_Q$ selected QSs, and the WF solution in each QS requires a POD-Galerkin model. In the global approach [50], the collected WF data from a selected set of QSs are used en bloc to produce a single set of POD modes, and thus the WFs in all the $N_Q$ selected QSs can be predicted by just one POD-Galerkin model. A preliminary study has demonstrated that the global model is superior to the individual model in 2D simulation of quantum dots (QDs) in terms of efficiency and accuracy [50]. In order to further understand the effectiveness and adaptability of the POD-Galerkin methodology, the global simulation approach is extended in this work to investigate 2D QD structures for two crucial applications. The first application involves external electric fields applied to a QD structure, which is of interest to applications of nano electronic and photonic devices. For the second application, the QD structures with internal potential variations are investigated with periodic BCs, which are useful for simulation and design of nanostructures and materials. Since Fourier modes are considered as a natural basis of periodic structures, the Fourier-based plane-wave (FPW) approach, similar to the method used in DFT, is also applied to a 2D periodic structure to further examine the effectiveness of the POD-Galerkin methodology.

## Method

The electron WF $\psi(\vec{r})$ in a quantum structure is determined by the Schrödinger equation,

$$\nabla \cdot \left[ -\frac{\hbar^2}{2m^*} \nabla \psi(\vec{r}) \right] + U(\vec{r})\psi(\vec{r}) = E\psi(\vec{r}), \tag{1}$$

where $\hbar$ is the reduced plank constant, $\vec{r}$ the position, $m^*$ the effective mass, $U(\vec{r})$ the potential energy, and $E$ the total energy of the electron. To reduce the DoF for the numerical WF solution, the Schrödinger equation is projected onto a mathematical space constituted by an optimal set of basis functions (or modes) $\eta(\vec{r})$. The POD process [13,14] is chosen to generate the optimal set of modes, which maximizes the mean square of the WF projection onto the mode over multiple sets of collected WF solution data,

$$\left\langle \left[ \frac{\psi \cdot \eta}{|\eta|} \right]^2 \right\rangle, \tag{2}$$

where the brackets $\langle \rangle$ denote the average over the WF data sets accounting for the parametric variations in the electric field/potential and BCs, and the inner product $\psi \cdot \eta$ and the L2 norm of $\eta$ are given below

$$\psi \cdot \eta = \int_\Omega \psi(\vec{r})\eta(\vec{r})d\Omega, \text{ and } |\eta| = \sqrt{\int_\Omega \eta(\vec{r})^2 d\Omega}. \tag{3}$$

This data-projection process ensures that the modes $\eta(\vec{r})$ contain the maximum least squares (LS) information of the system embedded in the collected WF data and leads to an eigenvalue problem for the spatial autocorrelation function $\boldsymbol{R}(\vec{r},\vec{r}')$ of the WF data,

$$\int_{\Omega'} \boldsymbol{R}(\vec{r},\vec{r}')\eta(\vec{r}')d\vec{r}' = \lambda\,\eta(\vec{r}), \tag{4}$$

where the eigenvalues $\lambda$ represent the mean squared WF information captured by each mode and $\boldsymbol{R}(\vec{r},\vec{r}')$ is given by

$$\boldsymbol{R}(\vec{r},\vec{r}') = \langle \psi(\vec{r}) \otimes \psi(\vec{r}') \rangle \tag{5}$$

with $\otimes$ as the tensor product. Using the POD modes generated from (4), the WF can be expressed via a linear combination of these modes,

$$\psi(\vec{r}) = \sum_{j=1}^{M} a_j \eta_j(\vec{r}), \tag{6}$$

with $M$ as the selected number of modes, where $M$ determines the DoF and the simulation efficiency and accuracy, and $a_j$ are weighting coefficients responding to the parametric variations in the simulation.

In a multi-dimension structure with a fine spatial resolution, the large dimension of $\boldsymbol{R}$ may be difficult to manage. The method of snapshots [51,52] is applied to convert the eigenvalue problem in (4) from a discrete space domain with a large dimension of the $N_r \times N_r$ to a sampling domain with a dimension of $N_s \times N_s$, where $N_r$ and $N_s$ are the numbers of spatial grid points and data samples/snapshots, respectively, and in general $N_s \ll N_r$. Instead of $N_r$ eigenvalues given in (4), the snapshot method solves only the first $N_s$ eigenvalues and POD modes. The number of data samples, $N_s$, thus needs to be large enough to ensure that $N_s > M$ and $\lambda_{N_s}$ is many orders smaller than $\lambda_1$ to minimize the numerical error resulting from the POD prediction since the theoretical LS error with an $M$-mode POD model is given as [7],

$$Err_{M,ls} = \sqrt{\sum_{i=M+1}^{N_s} \lambda_i \bigg/ \sum_{i=1}^{N_s} \lambda_i}, \tag{7}$$

where the eigenvalues are arranged in descending order. Theoretically, (7) is valid only if the data quality is sufficient; namely the parametric settings in the POD simulation fall within the bounds of the parametric variations used in the training and the data for the training is accurate enough. Also, differently from POD applications to many other fields, the global POD approach in this study collects WFs from all $N_Q$ selected QSs and the modes thus represent the solution for WFs in all $N_Q$ selected QSs. As a result, (7) offers a theoretical prediction of the LS error averaged over all selected WFs, as will be discussed later in the demonstrations.

To close the POD simulation methodology, the Schrödinger equation in (1) is projected onto the POD modes via the Galerkin projection. The projection along the $i$th POD mode $\eta_i$ is given as

$$\int_\Omega \nabla\eta_i(\vec{r}) \cdot \frac{\hbar^2}{2m^*} \nabla\psi(\vec{r})d\Omega + \int_\Omega \eta_i(\vec{r})U(\vec{r})\psi(\vec{r})d\Omega - \int_S \eta_i(\vec{r})\frac{\hbar^2}{2m^*}\nabla\psi(\vec{r})\cdot d\vec{S} = E\int_\Omega \eta_i(\vec{r})\psi(\vec{r})d\Omega. \quad (8)$$

Using (6) in (8), the quantum POD-Galerkin simulation methodology is thus represented by an $M \times M$ eigenvalue problem for $\vec{a} = [a_1, a_2, \ldots, a_j, \ldots, a_M]^T$,

$$\boldsymbol{H}_\eta \vec{a} = E\,\vec{a}, \quad (9)$$

where $\boldsymbol{H}_\eta$ is the Hamiltonian in the POD space denoted as

$$\boldsymbol{H}_\eta = \boldsymbol{T}_\eta + \boldsymbol{U}_\eta + \boldsymbol{B}_\eta \quad (10)$$

with the interior kinetic energy matrix expressed as

$$T_{\eta\,i,j} = \int_\Omega \nabla\eta_i(\vec{r}) \cdot \frac{\hbar^2}{2m^*}\nabla\eta_j(\vec{r})d\Omega, \quad (11)$$

the potential energy matrix expressed as

$$U_{\eta\,i,j} = \int_\Omega \eta_i(\vec{r})U(\vec{r})\psi(\vec{r})d\Omega, \quad (12)$$

and the boundary kinetic energy matrix given as

$$B_{\eta\,i,j} = -\int_S \eta_i(\vec{r})\frac{\hbar^2}{2m^*}\nabla\psi(\vec{r})\cdot d\vec{S}. \quad (13)$$

The $i$th eigenvector $\vec{a}_i$ in (9) corresponds to the $i$th QS and eigenenergy $E_i$, and the $j$th element of $\vec{a}_i$ in (6) represents the $j$th mode's weight for the $i$th QS. With homogeneous Neumann and Dirichlet BCs, the boundary kinetic energy matrix elements in (13) vanish. For a structure with periodic BCs, the surface normal vectors at the periodic boundary surfaces in (13) cancel out causing the boundary kinetic energy matrix to vanish as well.

Note that the formulations given in (8) - (11) apply to the projection of the Schrödinger equation onto any selected basis, including the Fourier plane-wave basis. The number of DoF needed to reach a desired accuracy in simulation depends on how well and concisely the selected basis functions portray the solution resulting from the potential variation and BCs.

## Demonstrations

In this work, the QDs are formed via the conduction band offset at the GaAs/InAs heterojunction; the material parameters in the simulations include effective masses $m^*_{GaAs} = 0.067m_o$ and $m^*_{InAs} = 0.023m_o$, and the band offset $\Delta E = 0.544\text{eV}$. Training data of WFs needed for POD mode generation are collected from DNSs using the finite difference method [53]. Two test structures are investigated below. In each structure, the training process for the QD structure is described first followed by the demonstrations.

**Quantum-dot structure subjected to external electric field.** This test case first performs the training of POD modes via a set of single-component electric fields from two orthogonal directions in $x$ and

*y*. The domain of the underlining nanostructure shown in Fig. 1 consists of a 4 × 4 grid of QDs, where each QD with a size of 4nm × 4nm is separated by 1nm from its adjacent QDs with a GaAs boundary spacer of 2.5 nm on each side of the domain. DNSs of the Schrödinger equation for this QD structure are carried out using a fine resolution of 241 ×241 (a grid size of 0.1 nm in either direction) with homogeneous Dirichlet and Neumann BCs to collect the WF data for the first 6 QSs. A fine mesh is needed to offer accurate eigenenergies and WFs of several nearly degenerate QSs in this QD structure. To account for field variations, in addition to a zero-bias simulation, DNSs of the QD structure are performed at 8 electric fields in each of the *x* and *y* directions with linearly spaced magnitudes between -35kV/cm and +35kV/cm. The sampled WF data in the QD structure collected from the 2 sets of orthogonal electric fields, together with the unbiased WF data, are combined to generate one set of POD modes that represent all 6 trained QSs. With 6 selected QSs for each field, a total of 102 samples/snapshots of the WF data is implemented in the method of snapshots [51,52] to solve the first 102 eigenvalues and POD modes from (4). This quantum POD-Galerkin model for $\vec{a}$ given in (9) is therefore constructed with the coefficients evaluated from (11)-(13) via its POD modes. After POD-Galerkin simulation using (9) to determine $\vec{a}$, post processing is needed to calculate the WF from (6).

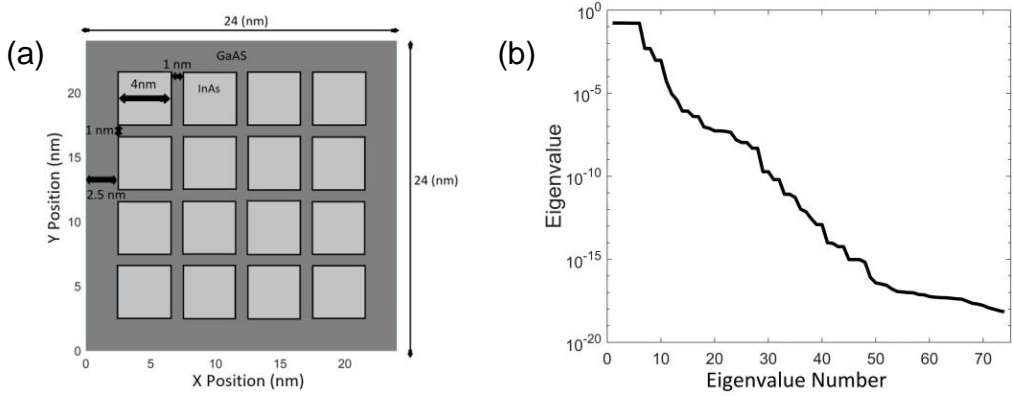

Figure 1. (a) Underlining GaAs/InAs QD structure for the demonstration of the POD-Galerkin simulation methodology. (b) Eigenvalue for the WF data in descending order.

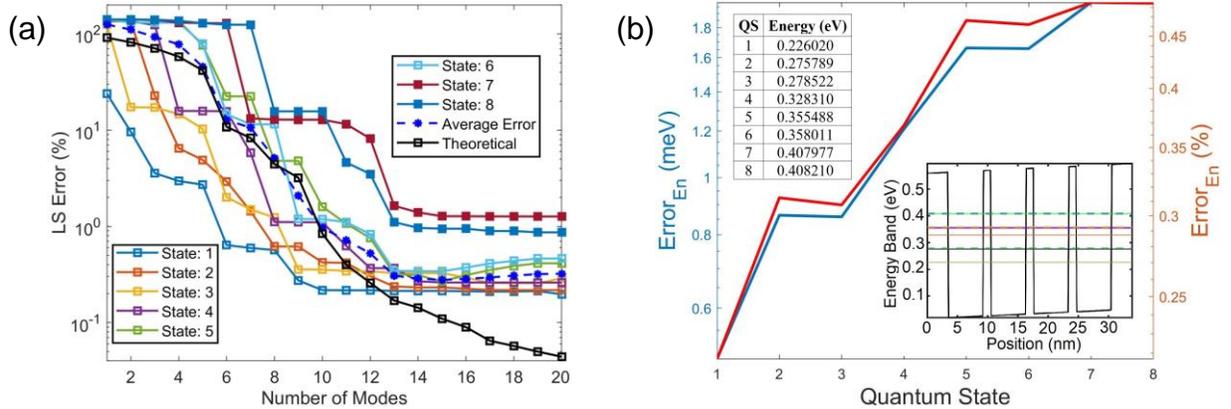

Figure 2. (a) POD LS errors for WFs in QSs 1-8. The average LS error (blue dashed line) is averaged over the 6 trained QSs, and the theoretical LS error $Err_{M,ls}$ is also for the first 6 QSs. (b) Absolute error of the estimated POD eigenenergy, where two insets include the eigenenergy obtained from the DNS, one in a table and the other in a band diagram along the diagonal direction of the QS structure from (0,0) to (24nm,24nm).

It is always informative to first observe the eigenvalues that offer information on the effectiveness of the trained POD modes. Eigenvalues of the first 6 modes shown in Fig. 1b vary very little, which reveals that the first 6 POD modes tend to maximize the information on all the 6 selected trained QSs. This is because the global quantum POD approach [50] implemented in this study generates POD modes that represent WFs in all $N_Q$ selected QSs, where $N_Q = 6$ in this case. Apparently, the most essential LS information is captured by the first $N_Q$ POD modes, and thus the eigenvalue decreases sharply beyond the $N_Q$-th mode. By Mode 11, the eigenvalue drops more than 3 orders of magnitude from the first mode and continues decreasing drastically beyond the 11th mode. As shown in Fig. 2a, the theoretical LS error estimated by (7) for the first 6 QSs becomes below 1% with 10 more modes. Moreover, one can see in Fig. 1b that after the 52nd mode the eigenvalue drops nearly 16 orders of magnitude from the first value and starts decreasing very slowly due to the limited computer precision.

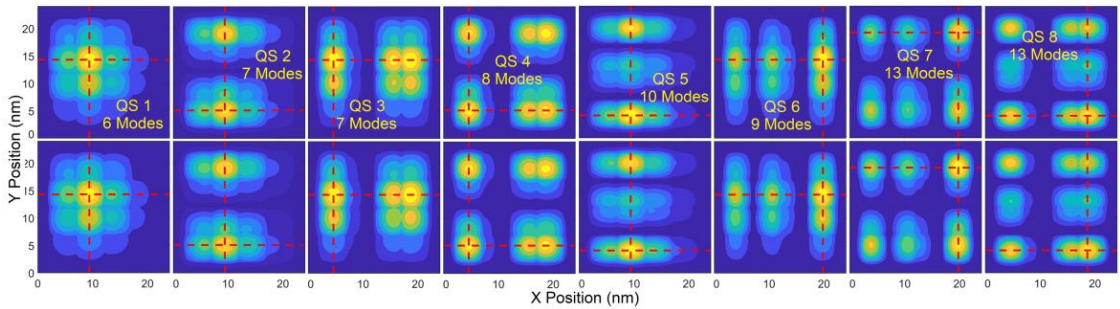

Figure 3. Contour of $|\psi|^2$ predicted by POD-Galerkin and DNS in upper and lower rows, respectively, in QSs 1-8. The horizontal and vertical red dashed lines reference the cross-sectional profiles of $|\psi|^2$ visualized in Fig. 4.

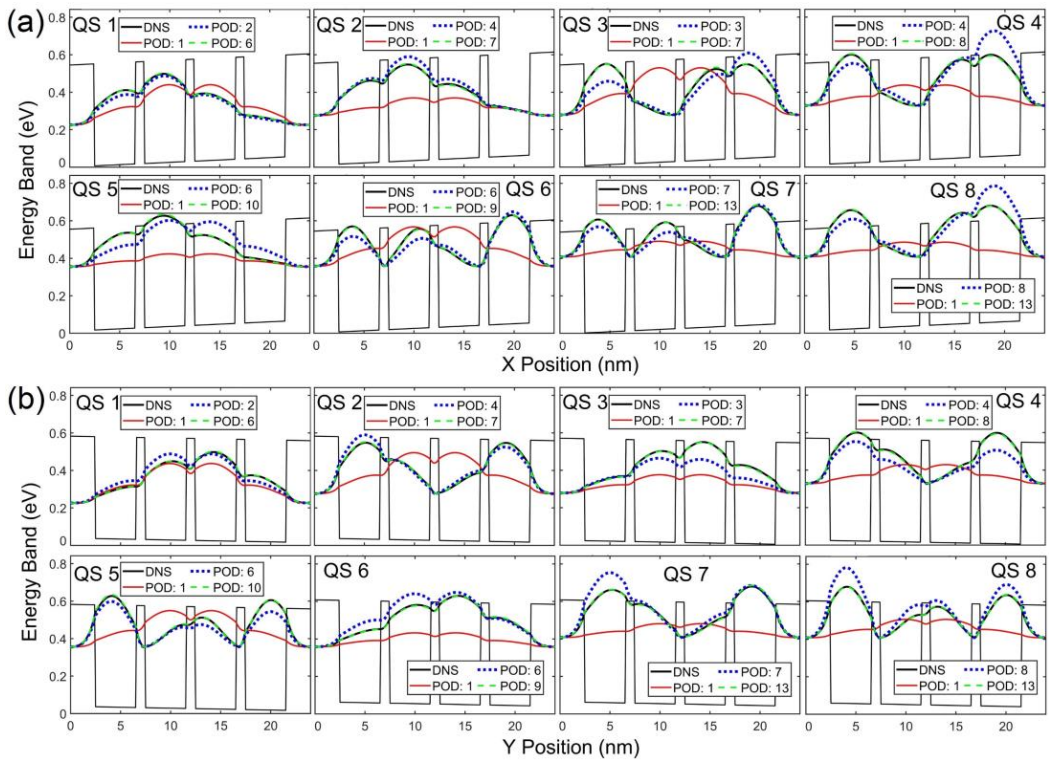

Figure 4. Profiles of $|\psi|^2$ in each state along (a) *x* and (b) *y* via the red dashed lines indicated in Fig. 3.

To test the validity of the POD-Galerkin methodology, a test electric field of $\vec{E} = (25\hat{x} - 10\hat{y})$ kV/cm is applied to the QD structure. The LS error of the POD WF in each state is first illustrated in Fig. 2a, compared to the theoretical LS error and the average LS error. Fig. 2a shows that the LS errors of WFs in all trained QSs predicted by the POD simulation are near or below 1% with 11 or more modes and their errors are all below 0.37% when incorporating 13 modes. The POD model is in general more effective in the lower QSs. As can been seen, the LS errors in QSs 1-4 are all near or below 1% when 8 modes are included, and POD WFs in QSs 1-3 reach an error near or below 0.6% and 0.42% with 9 and 10 modes, respectively. Even the POD WFs in the untrained 7th and 8th states reach an LS error near 1.6% and 1.1% with 13 modes and stay near 1.25% and 0.95% with more modes, respectively. The theoretical error $Err_{M,ls}$ in (7) predicts the average POD LS error for the first 6 QSs quite well until 13 modes where the average POD LS error is as low as 0.3%; this indicates a good data quality for this test case. Due to the numerical errors and computer precision, the average POD LS error starts deviating from $Err_{M,ls}$ beyond 13 modes and eventually converges to a range between 0.28% to 0.32%.

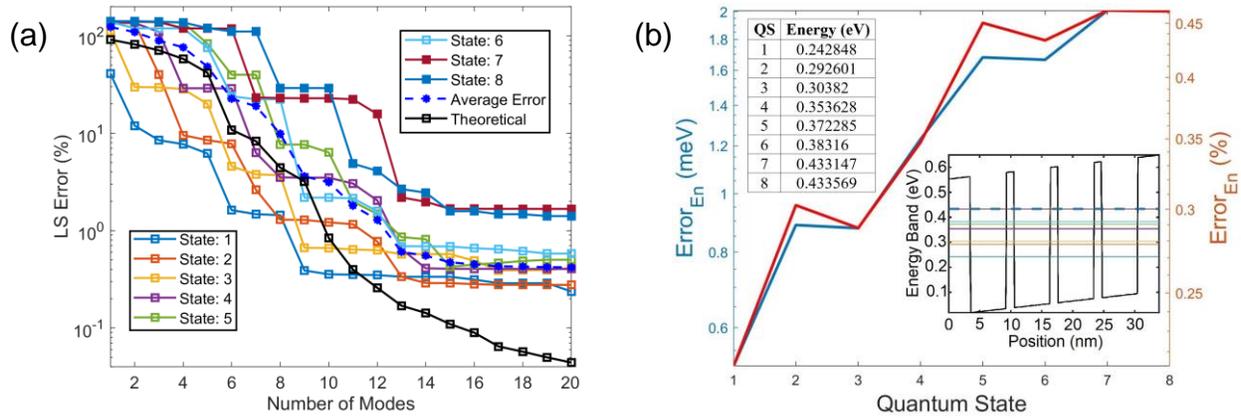

Figure 5. (a) POD LS errors for WFs in QSs 1-8. The average POD LS error is averaged over the 6 trained QSs, and the theoretical LS error $Err_{M,ls}$ is identical to that in Fig. 2. (b) Absolute error of the estimated POD eigenenergy with two insets of the eigenenergy obtained from the DNS, one in a table and the other in a band diagram along the diagonal direction of the QS structure from (0,0) to (24nm,24nm).

In addition to the accuracy of the predicted WFs, Fig. 2b illustrates the excellent agreement between the eigenenergies predicted by the POD-Galerkin methodology and by DNS of the Schrödinger equation, where the deviation for QSs 1-8 for the POD-Galerkin prediction ranges from 0.217% to 0.485%. It is interesting to observe the consistent increase in the deviation from the lower to higher QSs except for the nearly degenerate states, QSs 2-3, 5-6 and 7-8, where the eigenenergy differences in these degenerate states estimated from DNS are 2.733meV, 2.523meV and 0.233meV, respectively. The eigenenergy differences in these 3 sets of nearly degenerate states (2.728meV, 2.52meV, 0.231meV, respectively) predicted by the POD-Galerkin methodology are well preserved, including the untrained 7th and 8th QSs.

Contours and cross-sectional profiles of $|\psi|^2$ in each QS obtained from the POD and DNS are illustrated in Figs. 3 and 4, respective, where the maximum number of POD modes in each state is selected for its LS error near or below 1.5%. With a deviation near 1.5%, contours and profiles between these two approaches are nearly indistinguishable. The first mode of the POD-Galerkin solution always provides the mean of the training data. As seen in Fig. 4, the one-mode POD model offers the unbiased (symmetric)

$|\psi|^2$ in each QS since the training electric fields are symmetric about the zero field. Similar to Fig. 2, it is shown in Figs. 3 and 4 that 6-8 modes are needed in QSs 1-4 for the POD methodology to achieve a very good agreement (near or below a 1.5% LS error) with the DNS, while 9 or 10 modes are needed in the higher states (QSs 5 and 6). For the untrained 7th and 8th states POD WFs to reach a similar accuracy, 13 modes are needed. Figs. 2-4 clearly illustrate that the POD-Galerkin methodology is capable of extrapolating the WFs and the eigenenergies with a high accuracy even in the untrained 7th and 8th states if a few more modes are included.

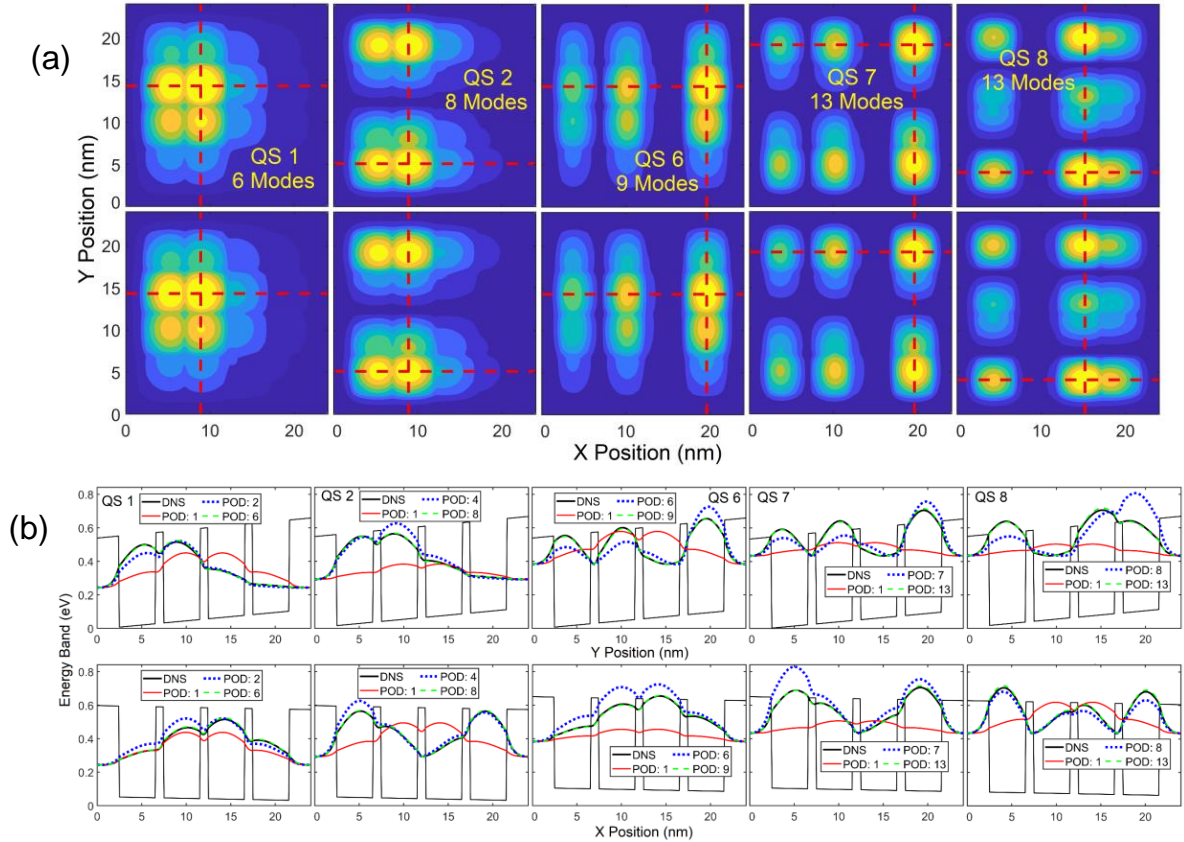

Figure 6. (a) Contours and (b) cross sectional profiles of $|\psi|^2$ in QSs 1, 2, 6, 7 and 8. In (a), the upper and lower rows show $|\psi|^2$ obtained from POD-Galerkin and DNS, respectively. The dashed horizontal and vertical lines in (a) reference the profiles visualized in (b) along *x* and *y* in upper and lower rows, respectively.

To further illustrate the extrapolation capability, an electric field of $\vec{E} = 50\,\hat{x} - 10\,\hat{y}$ (kV/cm) is applied to the QD structure. With the *x*-component field beyond the maximum training field (35k/cm), as expected in Fig. 5a, the error in each state declines more slowly and more modes are needed to reach a similar accuracy to the previous test case presented in Fig. 2. For example, 6, 7, 6, 8, 10, 9, 13 and 13 modes are needed from QSs 1 to 8 in sequence to reach an error below 2% in Fig. 2, while in Fig. 5 for this case beyond the training field, 6, 8, 9, 12, 11, 12, 14 and 15 modes are needed. Due to the applied field greater than the training fields, the data quality in this case is not as good as that in the previous case. As a result, the average LS error of the POD approach in Fig. 5 is slightly greater than that in Fig. 2 beyond 5 modes and does not follow the theoretical LS error (estimated based on the training data for the first 6 QSs) as close as the previous case. Nevertheless, compared to the interpolation case presented in Figs. 2-4, the POD-

Galerkin model still offers a very good prediction in the extrapolation case if a few more modes are included. Even in the untrained 7th and 8th QSs beyond the training fields, the POD LS error in QS 7 is as small as 2.2% and 1.67% with 13 and 15 modes, respectively, and 2.67% and 1.6% with 13 and 15 modes in QS 8, respectively. Excellent accuracy of the POD eigenenergy is also observed at this higher field, as displayed in Fig. 5b, where the deviation for QSs 1-8 for the POD prediction ranges from 0.213% to 0.463%. Unlike the lower field case, only QSs 7 and 8 are nearly degenerate in this case, whose eigenenergy difference estimated in DNS is as small as 0.422meV and is 0.419meV predicted by the POD-Galerkin model.

Contours and profiles of $|\psi|^2$ along $x$ and $y$ directions in several QSs predicted by the POD-Galerkin simulation and DNS are compared in Figs. 6a and 6b, where the maximum number of modes is chosen for the POD LS error near or below 2% in QSs 1, 2 and 6 and 2.5% in the untrained states 7 and 8. Accurate results for both WFs and eigenenergies presented in Fig. 5 and 6 highlight the capability of the POD-Galerkin methodology even in the extrapolation situations beyond both training fields and trained QSs, which is difficult to achieve in typical machine learning methods. Unlike most machine learning methods, the POD-Galerkin simulation methodology incorporates the first principles, as described in (8), by projecting the Schrödinger equation onto the POD modes. This projection offers a clear guideline to accurately predict the solution in response to the field variations even beyond the training fields in the untrained QSs. It is also worth noting that, even though the WF data were collected for electric fields in $x$ and $y$ directions separately, the POD modes generated from the combined data sets are able to accurately predict the WFs in each state subjected to an electric field in any direction.

**QD structure with internal potential variation and periodic BCs.** The next case extends the quantum POD-Galerkin methodology to a nanostructure with internal potential variation and periodic BCs. The underlining nanostructure of a 3×3 grid of GaAs/InAs QDs is shown in Fig. 7a. Each QS is 4nm×4nm in size adjacently separated by 1.5nm with a 1.25nm GaAs boundary spacer on each side. In addition, a potential energy profile of 5 pyramids each with a 4.8nm×4.8nm base shown in Fig. 7b is superposed on the band energy of the QD structure. The potential energy profile with variation of each pyramid height is chosen in this investigation to demonstrate the proposed physics-informed learning algorithm derived from the first principles to predict the WFs in response to the variation of the internal potential profile with periodic BCs, which appear in many applications in quantum nanostructures and materials, including DFT simulations [36,45,47-49].

To account for variation of the internal potential in the WF data collection, DNS of the QD structure described in Figs. 7a and 7b is first performed with only one of the five pyramids at a time. Five different heights for each pyramid potential energy varying equally from 0.07 eV to 0.35eV are included in DNSs. When using only these 25 pyramid potential samples to train the POD modes, it was found that the LS error of the POD-Galerkin model is relatively large, as expected due to the poor data quality. More specifically, the influences among the five pyramids are not included in the collected data. To thoroughly account for the influences among different heights of the pyramid potentials, if five different heights are selected for each pyramid, a combination of all different heights for the five pyramids should be included in the DNSs to collect the WF data. This would lead to an enormous number of additional samples, i.e., $5^5 = 3125$, which requires an immense computational effort to collect the WF data from DNSs and evaluate the POD

Hamiltonian elements in (11)-(13). To minimize the training effort, five pyramids varying together with the same (five) heights from 0.07 eV to 0.35eV in DNS are used to collect just one additional sample of WF data with the hope that the physical principles enforced by the Galerkin projection would intelligently predict the influences among different heights of pyramids. As will be seen later, this setting actually works reasonably well.

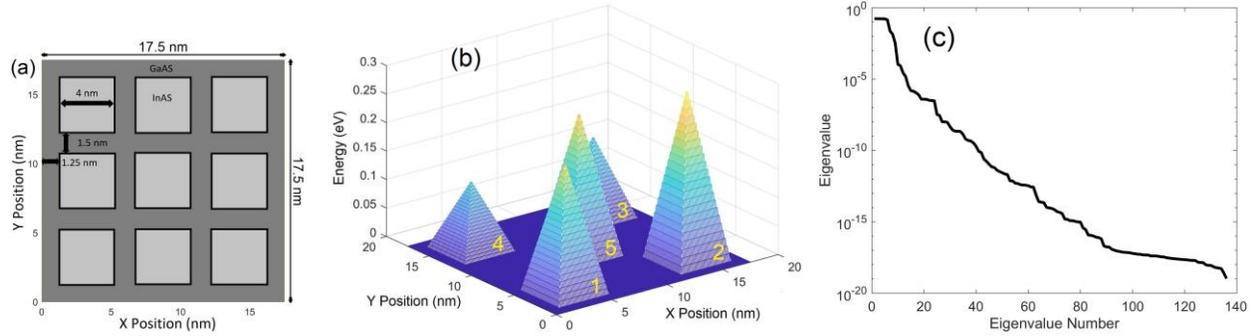

Figure 7. (a) GaAs/InAs QD structure with periodic BCs. (b) Potential energy with a profile of 5 pyramids applied to the QD structure in (a). (c) Eigenvalue of WF data in descending order.

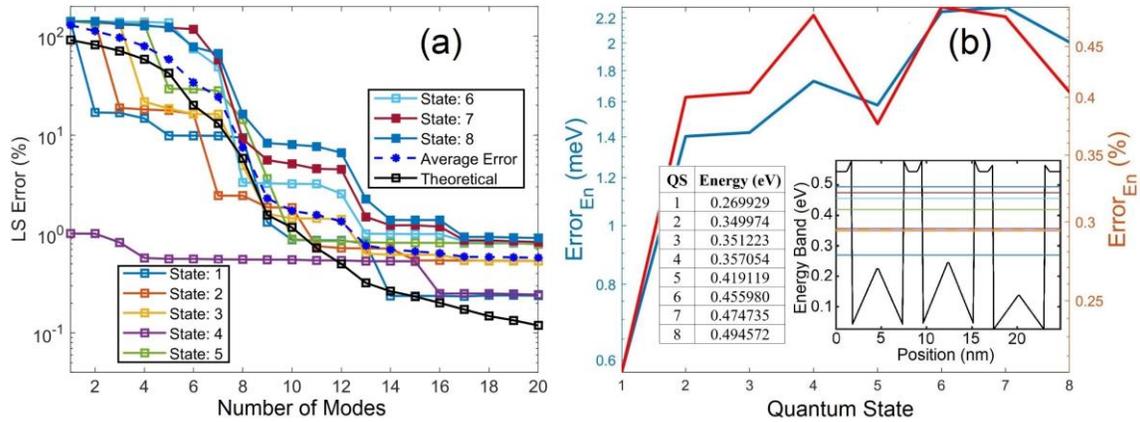

Figure 8. (a) POD LS errors for WFs in QSs 1-8. The average LS error is averaged over the 6 trained QSs, and the theoretical LS error $Err_{M,ls}$ is also for the first 6 QSs. (b) Absolute error of the estimated POD eigenenergy with two insets including the eigenenergy obtained from the DNS, one in a table and the other in a band diagram along the diagonal of the QS structure in Fig. 7 from (0,0) to (17.5nm, 17.5nm).

In each simulation, only the WFs in the first 6 ($N_Q$) QSs are collected. Using data collected from the QD structure with variations of five pyramid potentials, together with the structure without any pyramid, there are in total 186 samples/snapshots (i.e., (6×5 +1)×6) of WF data collected to solve POD modes and eigenvalues from (4) via the method of snapshots [51,52]. Similar to Fig. 1b, the eigenvalues displayed in Fig. 7c remain nearly unchanged for the first $N_Q$ modes, and decrease sharply beyond the $N_Q$-th mode. The eigenvalue also reveals a more than 3-order drop from the first to the 11 modes and decreases considerably more slowly after dropping 16 orders of magnitude from the first mode due to the computer precision.

In this demonstration, the heights of Pyramids 1 to 5, as labeled in Fig. 7b, are randomly selected as 0.23eV, 0.3eV, 0.14eV, 0.12eV and 0.25eV in sequence. The average LS error of the WF predicted by the POD-Galerkin model shown in Fig. 8a for the first 6 QSs agrees quite well with the theoretical LS error estimated in (7) until 11 modes although not as well as that in Fig. 2a. This indicates that the data quality associated with Fig. 2a is better than that for this test case. However, the simple training in this case to account for influences among the pyramids with unequal heights still offers very accurate prediction of WFs (near 1% LS error) in the trained (first 6) states with just 12 to 13 modes. Because of a better data quality, the POD-Galerkin model for the external field case in Fig. 2b is more effective than the model for the periodic BC case. For example, to reach an LS error near 1% for all trained states, 10 or 11 modes are needed in Fig. 2a while 13 modes in Fig. 8a. When using 13 or more modes, the average LS error near 0.3% is observed in Fig. 2a while 0.78% in Fig. 8a. Also note that the LS errors of the untrained 7th and 8th states in this periodic BC case are as small as 1.5% and 2.25% with 13 modes, respectively, reduce to 1.2% and 1.4% with 14 modes, and below 1% beyond 16 modes. The eigenenergy predicted by the POD-Galerkin simulation in Fig. 8b is as accurate as that in Fig. 2b, and QSs 2 and 3 are nearly degenerate with an eigenenergy difference of 1.249meV calculated in DNS, as shown in the insets. The POD-Galerkin model however predicts a difference of 1.269meV.

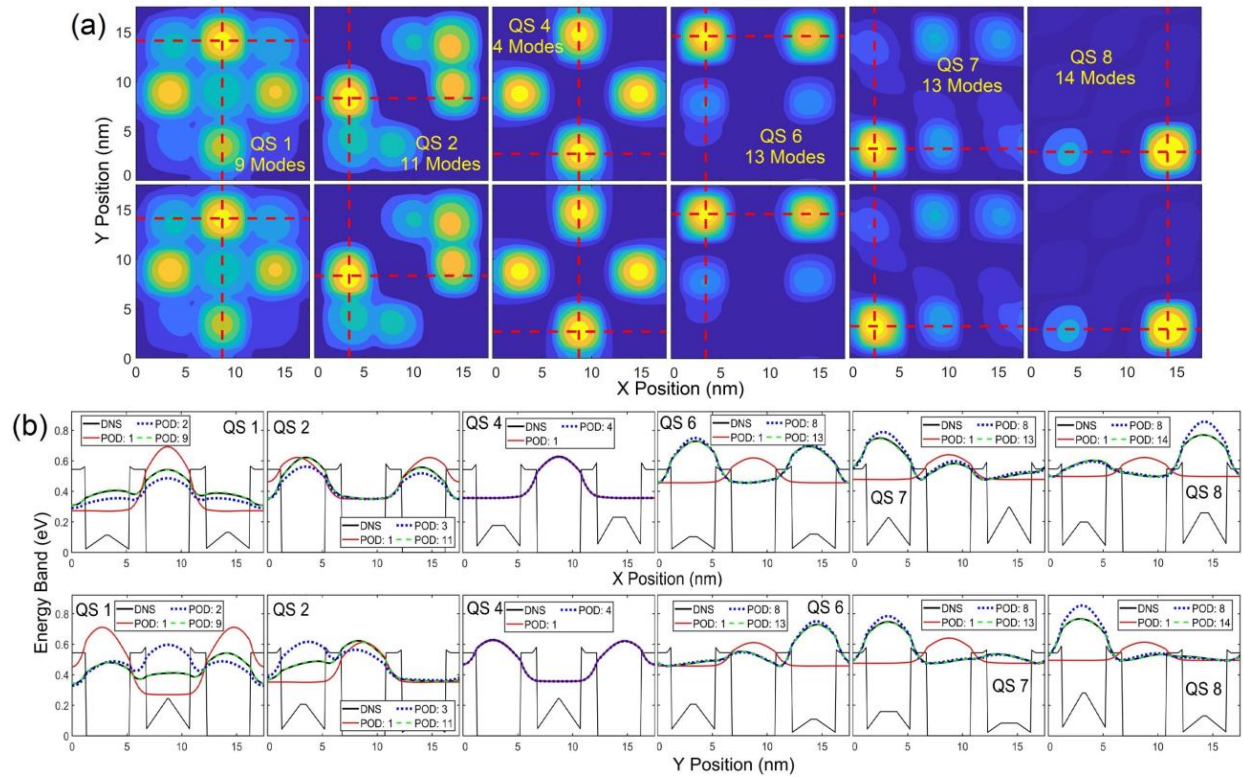

Figure 9. (a) Contours and (b) cross sectional profiles of $|\psi|^2$ in QSs 1, 2, 4, 6, 7 and 8. In (a), the upper and lower rows show $|\psi|^2$ obtained from POD-Galerkin and DNS, respectively. The dashed horizontal and vertical lines in (a) reference the profiles visualized in (b) along *x* and *y* in the upper and lower rows, respectively.

Contours and profiles of $|\psi|^2$ are illustrated in Figs. 9a and 9b, respectively, for QSs 1, 2, 4, 6 7 and 8, where the maximum number of POD modes in each state is selected for the LS error below 1.5%. Similar to results presented in Figs. 2-4, to reach a high accuracy, more modes are needed when the eigenenergy is closer to the QD barrier energy. Again, the result of the first mode represents the mean of the training WF data which are symmetric about the center of the QD structure, as revealed in the one mode POD WF solution for all states. In this demonstration, the WF in QS 4 appears to be nearly symmetric and thus can be well represented by the one-mode POD solution (see the QS 4 profiles in Fig. 9b along both *x* and *y* directions) with just a 1% LS error, as shown in Fig. 8a. The QS-4 LS error decreases to 0.55% between 4 and 15 modes and drops to 0.24% beyond 15 modes. This indicates that the QS-4 WF is not perfectly symmetric, and 4 or more modes are needed if higher accuracy is desired. For the untrained 7th and 8th states' POD WFs, high accuracy can still be achieved when using 13 or 14 modes.

For periodic functions, the Fourier basis is usually assumed to approximate the solution of the problem. An additional demonstration is therefore illustrated using FPWs as the basis functions in (8) and (9) to derive an FPW model and to perform simulation of the periodic structure given in Fig. 7a with the pyramid potential variation in Fig. 7b. The LS errors against the DNS are shown in Fig. 10, where even with periodic BCs a large number of modes of FPWs is still needed to reach a reasonably small error due to complicated internal pyramid potential variation. For example, when using 40 modes, the minimum error is near 5.4% in QS 1 and the maximum is near 11% in QS 7. Using 225 modes, the minimum error reduces to 2.78% in QS 1 and the maximum is near 5.51% in QS 3. Beyond 225 modes, the LS error induced by the FPW approach continues decreasing very slowly. Nevertheless, this demonstration further validates the effectiveness of the POD-Galerkin methodology in which use of 13 – 14 modes for this periodic structure offers an LS error of WFs 5 to 11 times smaller in each of all the trained QSs than the FPW approach using 225 modes. For the untrained 7th and 8th states with 13-14 modes, POD-Galerkin leads to an LS errors 2 to 3.5 times smaller than the FPW approach using 225 modes. The error for the eigenstate energy predicted by the FPW approach given in Fig. 10 in each state is approximately 4.5 to 7 times as large as that in the POD-Galerkin model given in Fig. 8b.

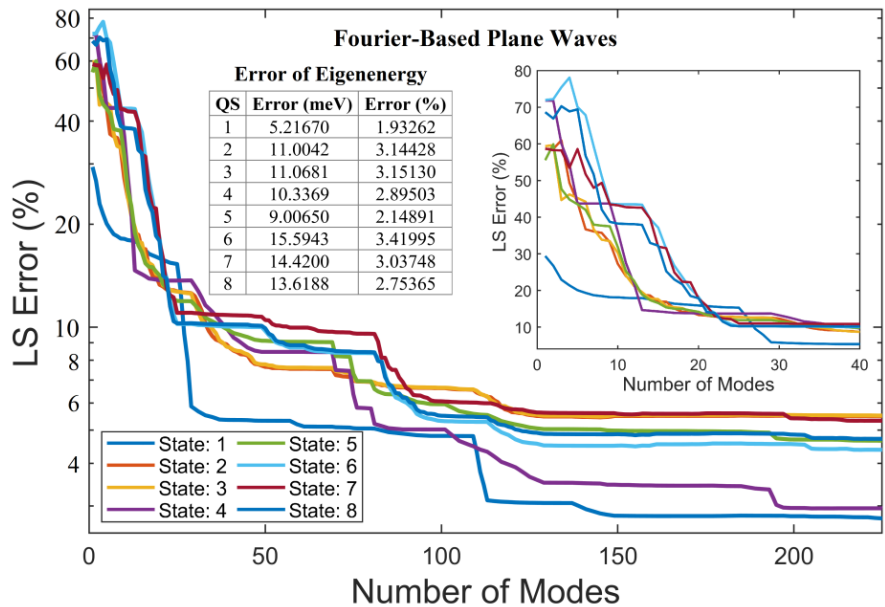

Figure 10. LS errors of WFs in QSs 1-8 for the first 225 modes, resulting from the FPW approach. The insets include a closer look at the LS errors for the first 40 modes and a table of the absolute error of the eigenenergy estimated from the plane-wave approach. The eigenenergy obtained from DNS is given in an inset of Fig. 8b.

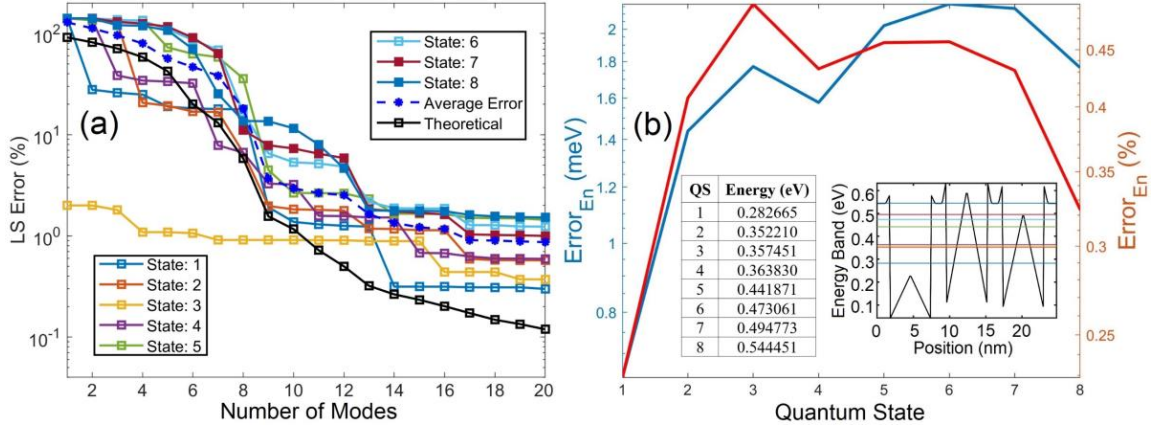

Figure 11. (a) POD LS errors for WFs in QSs 1-8. The average LS error is averaged over the 6 trained QSs, and the theoretical LS error $Err_{M,ls}$ is also for the first 6 QSs. (b) Absolute error of the estimated POD eigenenergy with two insets including the eigenenergy obtained from the DNS, one in a table and the other in a band diagram along the diagonal of the QS structure in Fig. 7 from (0,0) to (17.5nm, 17.5nm).

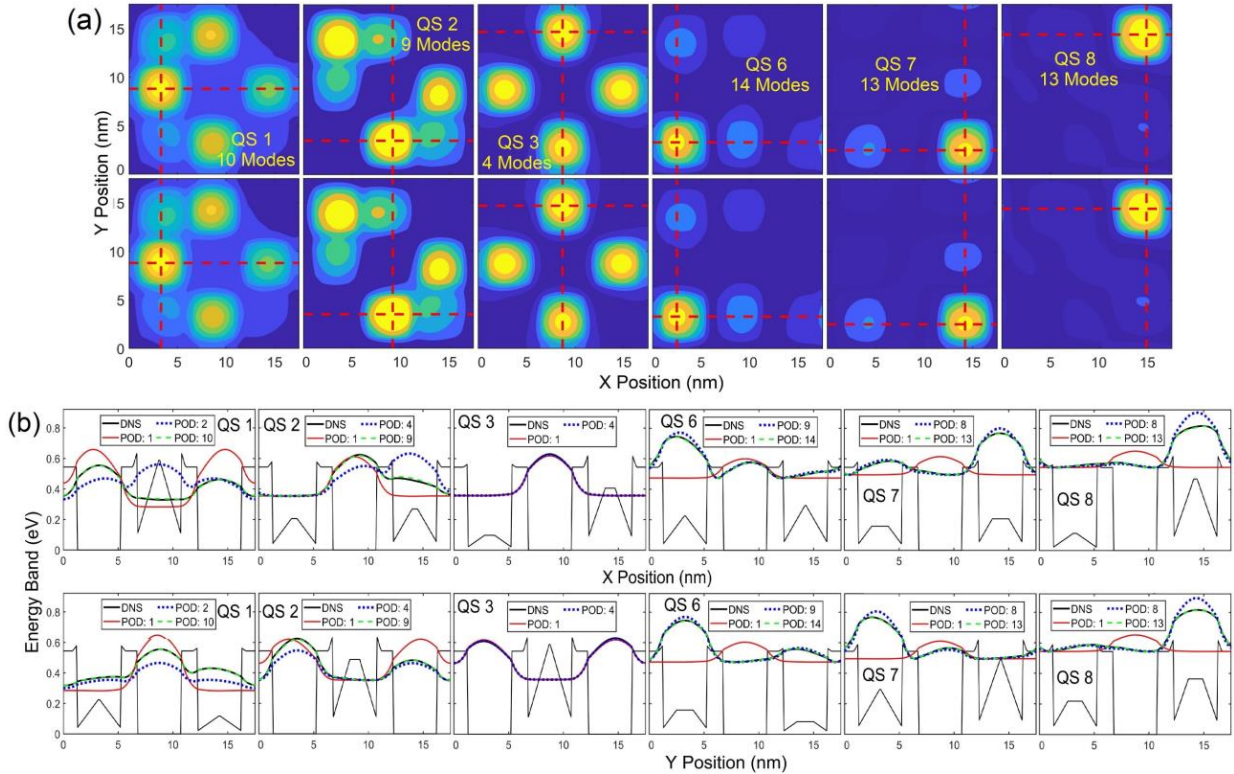

Figure 12. (a) Contours and (b) cross sectional profiles of $|\psi|^2$ in QSs 1, 2, 3, 6, 7 and 8. In (a), the upper and lower rows show $|\psi|^2$ obtained from POD-Galerkin and DNS, respectively. The dashed horizontal and vertical lines in (a) reference the profiles visualized in (b) along $x$ and $y$ in the upper and lower rows, respectively.

To further validate the extrapolation ability of the POD-Galerkin methodology for the QD structure with pyramid potential variation, Pyramids 1-5 are selected as 0.23eV, 0.3eV, 0.5eV, 0.12eV and 0.6eV in sequence, where heights of Pyramids 3 and 5 are evidently higher than the maximum training height (0.35eV). For this demonstration, the average LS error shown in Fig. 11a becomes slightly larger than that in Fig. 8a and moves away from the theoretical LS error. Instead of 9 or 10 modes to reach an average error near 2% in Fig. 8a, 13 modes are needed for this extrapolation case. Nevertheless, the POD-Galerkin methodology still offers an accurate prediction of the WFs in this case with a small number of DoF, especially in the lower QSs. For example, the LS error is all below 2% with 9 modes in QSs 1-3, 1.2% in QSs 1-4 with 15 modes and 0.6% in QSs 1-4 when using 17 modes. It is interesting to observe that somehow the untrained 7th QS reaches an error smaller than that in the 5th and 6th states with 13 or more modes. The LS error in the higher states (QSs 5-8) is all lower than 2% beyond 13 modes. The error of the POD eigenenergy in this case with higher pyramid potential heights shown in Fig. 11b is very similar to what was observed in Fig. 8b. Due to the larger variation of internal pyramid potential in this periodic structure, accuracy of the FPW approach further deteriorates and becomes considerably worse than the POD-Galerkin model. The FPW approach is therefore not presented for this case.

Figs. 12a and 11b illustrated the contours and profiles of $|\psi|^2$, respectively, for QSs 1, 2, 3, 6 7 and 8, where the maximum number of POD modes in each state is selected for the LS error below 2%. The contours and profiles of $|\psi|^2$ derived from the POD-Gakerlin prediction and DNS are nearly identical. Instead of QS 4 in Fig. 9, the QS-3 WF in this case is nearly symmetric, as shown in Fig. 12; as a result, the one-mode POD WF in QS 3 leads to a 2% LS error. More modes are needed to compensate for the non-symmetric portion of the WF, and the LS error reduces to 1.16% with 4 modes, 0.91% with 6 modes and 0.43% with 16 mods.

## Discussions

In summary, the thorough investigation of the quantum POD-Galerkin simulation methodology for two distinct QD structures has demonstrated its remarkable learning ability, which includes:

- **Extrapolation:** In both structures, the POD-Galerkin methodology predicts WFs and eigenenergies accurately with a small number of DoF even in the untrained QSs influenced by applied electric field or internal pyramid potential beyond the training conditions.
- **Simple training to account for effects of neighboring potential variations:** Using a very simple training scheme in which one varies the pyramid potentials all together to add only one additional snapshot of WFs (instead of several thousand snapshots to include all possible variations), the POD-Galerkin model is able to account for effects of the neighboring pyramid potentials with different heights to reach high accuracy, as illustrated in Figs. 8, 9, 11 and 12. This also works reasonably well even beyond the training pyramid potential height.
- **Single orthogonal field training:** As demonstrated in Figs. 1-6, POD modes trained by single orthogonal field components, together with the Galerkin projection, are able to accurately predict the WFs subjected to a field in an arbitrary direction. An accurate prediction can also be achieved in the untrained QSs and for electric field reasonably beyond the training field.

POD finds its modes by maximizing the mean square of the sampled data projection onto each orthogonal mode, as described in (2). This decomposition ensures that the first few modes contain essential LS information embedded in the WF data and thus only a small number of DoF is needed to reach a minimum LS error. Such an optimal set of POD basis functions is very different from the FPW basis that, although a natural choice for a periodic structure, may not be optimal for all periodic structures. As demonstrated in Fig. 10, due to complicated variation of interior pyramid potential, a large number of modes is needed for the FPW method to reach a reasonable agreement with the DNS. The POD-Galerkin learning ability also stems from the fact that the projection of the Schrödinger equation onto the trained POD modes offers a physics-informed compact model guided by the first principles to intelligently perform sophisticate interpolation and extrapolation. The physics-informed learning approach is also able to clearly distinguish the nearly degenerate states with energy difference below 0.25meV even in the untrained states beyond the training range. Our study reveals that the post-processing calculation of WFs in physical space in (6) is computationally considerably more intensive than the simulation in POD space to solve $\vec{a}$ in (9). The accurate prediction of the nearly degenerate states requires small grid size for DNSs in the training, which makes the post process for WFs in (6) even more time consuming.

Overall, the POD-Galerkin simulation methodology for 2D nanostructures offers a reduction in DoF by more than 3 orders of magnitude than DNS to achieve high accuracy, which leads to a computational speed-up of more than 2 orders. In addition, the POD-Galerkin methodology offers predictions of WFs and eigenenergies significantly more accurate than the FPW approach with a more than one-order reduction in DoF that leads to a reduction in computational time by one order of magnitude. Although high spatial resolution in DNSs is needed in the training to enhance the accuracy of the POD-Galerkin prediction, practically such high spatial resolution in WFs is not required for many applications that need the WFs. The POD-Galerkin simulation speed can then be further improved if the post process in (6) is performed with lower spatial resolution.

## Conclusion

The physics-informed POD-Galerkin simulation methodology based on the global quantum POD model [50] has been applied to investigate its learning ability for simulations of two QD structures, one influenced by external electric field and the other by internal potential and BCs. The extraordinary learning ability of the POD-Galerkin methodology has been demonstrated even in the untrained states subjected to applied field and internal potential beyond the training conditions. Such an effective learning results from the compact set of basis functions optimized by (2) and the Galerkin projection performed in (8) that empowers POD-Galerkin simulation to comply with the first principles. One important factor affecting the accuracy of the simulation methodology is the quality of the training data. In addition to the numerical accuracy for the data collected from the DNSs of the nanostructure, an appropriate range of parametric variations is needed, which should cover the POD-Galerkin simulation setting as much as possible. In case that the POD-Galerkin simulation setting is slightly outside the bounds of training parametric variations, a few more modes are needed to reach a good prediction. Overall, the POD-Galerkin simulation of 2D nanostructures provides a reduction in DoF by more than 3 orders of magnitude and computational time by 2 orders. Compared to the FPW approach, a one-order reduction in both DoF and computational time can be accomplished with considerably more accurate predictions of WFs and eigenenrgies. The proposed approach potentially can be applied to replace the most computationally intensive steps in self-consistent field calculations in DFT simulation, including the plane-wave basis function generation, Hamiltonian generation and eigenmode solver [54].

# Acknowledgements

This work was supported by the National Science Foundation, USA, under Grant OAC-2118079 and Grant OAC-1852102

# References

1. C. P. Moreno, P. J. Seiler, G. J. Balas, Model Reduction for Aeroservoelastic Systems. J. Aircraft, **51**, 280–290 (2014).
2. Z. Luo, G. Chen, Proper Orthogonal Decomposition Methods for Partial Differential Equations. SIAM Review, **63**, 231-245 (2021)
3. Z. Li, Y. Ma, L. Cao, and H. Wu, Proper orthogonal decomposition based online power-distribution reconstruction method. Annals of Nuclear Energy, **131**, 417–424 (2019).
4. J. Oliver, M. Caicedo, A. E. Huespe, J. A. Hernández, and E. Roubin, Reduced order modeling strategies for computational multiscale fracture. Computer Methods Appl. Mech. Eng., **313**. 560–595 (2017).
5. X. Sun, B. Li, X. Ma, Y. Pan, S. Yang, and W. Huang, Proper orthogonal decomposition-based method for predicting flow and heat transfer of oil and water in reservoir. J, Energy Resources Tech., **142**, 0124011-1240110 (2019).
6. W. Jia, B. T. Helenbrook, M.C. Cheng, Thermal modeling of multi-fin field effect transistor structure using proper orthogonal decomposition. IEEE Tran. Electron Devices, **61**, 2752–2759 (2014).
7. M.C. Cheng, "A reduced-order representation of the Schrödinger equation," AIP Advances, **6**, 095121, (2016).
8. L. De Lathauwer, B. De Moor, J. Vandewalle, A multilinear singular value decomposition. SIAM J. Matrix Analysis and Applications, **21**, 1253-1278 (2000).
9. M.E. Wall, A. Rechtsteiner, LM. Rocha, Singular value decomposition and principal component analysis. A practical approach to microarray data analysis, Springer, Boston, MA, 91-109 (2003).
10. K. Zhang, W. Jia, J. Koplowitz, P. Marzocca, M.-C. Cheng, Modeling of photovoltaic cells and arrays based on singular value decomposition, Semi. Sci. Tech., **28**, 035002 (2013).
11. H. Abdi, L.J. Williams, Principal component analysis. Wiley interdisciplinary reviews: computational statistics, **2**, 433-459 (2010).
12. M.E. Tipping, C.M. Bishop, Probabilistic principal component analysis. J. Royal Statistical Society: Series B (Statistical Methodology), **61**, 611-622 (1999).
13. J. L. Lumley, "The Structure of Inhomogeneous Turbulence," Atmospheric Turbulence and Wave Propagation, A. M. Yaglom and V.I. Tartarski, Ed. Moscow, 166-178 (1967).
14. J. L. Lumley, *Stochastic Tools in Turbulence.* Mineola, NY, USA: 1970; reprint, Dover publisher, 2007.
15. Radu, A., Duque, C.A. Neural network approaches for solving Schrödinger equation in arbitrary quantum wells. *Sci Rep* **12**, 2535 (2022).
16. S. Iserte, A. Macías, R. Martínez-Cuenca, S. Chiva, R. Paredes, E. S. Quintana-Ortí, Accelerating urban scale simulations leveraging local spatial 3D structure. J. Computational Science, **62**, 101741 (2022),
17. R. Abadía-Heredia, et al., A predictive hybrid reduced order model based on proper orthogonal decomposition combined with deep learning architectures, Expert Systems with Applications, **187**, 115910 (2022).
18. S. Fresca, A. Manzoni, POD-DL-ROM: Enhancing deep learning-based reduced order models for nonlinear parametrized PDEs by proper orthogonal decomposition, Comp. Methods Appl. Mech. & Eng., **388**, 114181 (2022).
19. E. Iuliano, D. Quagliarella, Proper orthogonal decomposition, surrogate modelling and evolutionary optimization in aerodynamic design, Computers & Fluids, **84**, 327-350 (2013).

20. D. Coenen, H. Oprins, I. De Wolf, Circuit-level thermal modelling of silicon photonic transceiver array using machine learning. 21st IEEE Int. Conf Thermal & Thermomech. Phenomena Electronic Systems (iTherm), 1-8, (2022).
21. D. Ahlman, F. Soderlund, J. Jackson, A. Kurdila and W. Shyy. Proper orthogonal decomposition for time-dependent lid-driven cavity flows, Numer. Heat Transfer, Part B. **42**, 285–306 (2002).
22. K. Lu, et. al, Review for order reduction based on proper orthogonal decomposition and outlooks of applications in mechanical systems. Mech. Sys. Signal Proce., **123**, 264–297 (2019).
23. D. Binion and X. Chen. 2017. A Krylov enhanced proper orthogonal decomposition method for frequency domain model reduction. Eng. Computations, vol. **34**, 285–306 (2017).
24. J. Faist et al., Quantum Cascade Laser, Science, **264**, 553-556 (1994).
25. H. Dakhlaoui, J.A. Vinasco, C.A. Duque, External fields controlling the nonlinear optical properties of quantum cascade laser based on staircase-like quantum wells, Superlattices and Microstructures, **155**, 106885 (2021).
26. S.A. Veldhuis, et al., Perovskite materials for light-emitting diodes and lasers, Adv. Mater.. **28**, 6804-6834 (2016).
27. J. Song, et al., Quantum dot light-emitting diodes based on inorganic Perovskite cesium lead halides (CsPbX3). Adv. Mater.. **27**, 7162-7167 (2015).
28. K.C. Kim, et al., Improved electroluminescence on nonpolar m -plane InGaN/GaN quantum wells LEDs. phys. stat. sol. (RRL), **1**, 125-127 (2007).
29. A. Pescaglini, et al., Three-Dimensional Self-Assembled Columnar Arrays of AlInP Quantum Wires for Polarized Micrometer-Sized Amber Light Emitting Diodes, ACS Photonics, **5**, 1318-1325 (2018).
30. X. Lan et al., Passivation using molecular halides increases quantum dot solar cell performance. Adv. Mater, **28,** 299-304 (2016)..
31. J. Pan et al., Automated Synthesis of Photovoltaic-Quality Colloidal Quantum Dots Using Separate Nucleation and Growth Stages. ACS Nano. **7,** 10158-10166 (2013).
32. F. Pelayo et al., Semiconductor quantum dots: Technological progress and future challenges. Science, **373**, eaaz8541 (2021).
33. F. Hetsch, N. Zhao, S. V. Kershaw, A. L. Rogach, Quantum dot field effect transistors. Materials Today, **16**, 312-325 (2013).
34. J. Wen, H. Hu, G. Wen, S. Wang, Z. Sun, S. Ye, Thin film transistors integrating $CsPbBr_3$ quantum dots for optoelectronic memory application. J. Physics D: Applied Physics, **54**, 114002 (2021).
35. M. Hu, Z. Yang, W. Zhou, A. Li, J. Pan, F. Ouyang, Field effect transistors based on phosphorene nanoribbon with selective edge-adsorption: A first-principles study. Physica E: Low-dimensional Systems and Nanostructures, **98**, 60-65 (2018).
36. K. Pathakoti, M. Manubolu, H.-M. Hwang, Nanostructures: Current uses and future applications in food science," J. Food and Drug Analysis, **25**, 245–253 (2017).
37. G. Cao, Applications of nanomaterials. *Nanostructures and Nanomaterials - Synthesis, Properties and Applications*, London, U.K. Imperial College Press, 391-418 2014).
38. Y. Zhang et al., Improved hetero-interface passivation by microcrystalline silicon oxide emitter in silicon heterojunction solar cells. Science Bulletin, **61**, 787–793 (2016).
39. M. H. N. Assadi, et al., Theoretical study on copper's energetics and magnetism in TiO2 polymorphs, J. Applied Phys., **113**, 233913–5 (2013).
40. F. F. Rastegar, N. L. Hadipour, H. Soleymanabadi, Theoretical investigation on the selective detection of SO2 molecule by AlN nanosheets. *J. Molecular Modeling*, **20**, 2439 (2014).
41. O. Brea, H. Daver, J. Rebek, F. Himo, Mechanism(s) of thermal decomposition of N-Nitrosoamides: A density functional theory study. *Tetrahedron*, **75**, 929-935 (2019).
42. T. Vo-Dinh, A. M. Fales, G. D. Griffin, et. al, Plasmonic nanoprobes: from chemical sensing to medical diagnostics and therapy. Nanoscale, **5**, 10127-10140 (2013).
43. M. F. Ng, M. B. Sullivan, S. W. Tong, P. Wu , First-principles study of silicon nanowire approaching the bulk limit. *Nano Letters* **11**, 4794-4799 (2011).

44. J. X. Mu, et. Al., Design, Synthesis, DFT study and antifungal Activity of Pyrazolecarboxamide Derivatives. Molecules, **68**, 26760990 (2016).
45. S. Chahib, D. Fasquelle, G. Leroy, "Density functional theory study of structural, electronic and optical properties of cobalt-doped BaSnO3. Materials Science in Semiconductor Processing, **137**, 106220 (2022).
46. S. Datta, Nanoscale device modeling: the Green's function method, Superlattices and Microstructures. **28**, 253-278 (2000).
47. E. Akhoundi, M. Houssa, A. Afzalian, The impact of electron phonon scattering on transport properties of topological insulators: A first principles quantum transport study, *Solid-St. Electronics*, **201**, 108587 (2023).
48. J. Cao, G. Gandus, T. Agarwal, M. Luisier, Y. Lee, Dynamics of van der Waals charge qubit in two-dimensional bilayer materials: Ab initio quantum transport and qubit measurement, Phys. Rev. Research, **4**, 043073 (2022).
49. M. Shin, S. Jeon, K. Joo, Efficient atomistic simulations of lateral heterostructure devices with metal contacts. Solid-St. Electronics, **198**, 108456 (2022).
50. M. Veresko, M. C. Cheng, An effective simulation methodology of quantum nanostructures based on model order reduction. Int. Conf. Simul. Semicond. Processes and Devices (SISPAD 2021), 64-68 (2021).
51. L. Sirovich, Turbulence and the dynamics of coherent structures. I–Coherent structures. II–Symmetries and transformations. III–Dynamics and scaling. Quart. Appl. Math., **45**, 561–571 and 573–590, 1987.
52. W. Jia, M. C. Cheng, A methodology for thermal simulation of interconnects enabled by model reduction with material property variation. J. Computational Sci., **61**, 101665 (2022).
53. Numerical approaches for solving the Schrödinger equation and implementing the quantum POD-Galerkin methodology in quantum dot structures can be found in the Matlab codes posted at https://github.com/CompResearchLab/POD-Schrodinger-Equation-Solver.
54. E. Di Napoli, S. Blügel, P. Bientinesi, Correlations in sequences of generalized eigenproblems arising in Density Functional Theory. Computer Physics Comm., **183**, 8, 1674-1682 (2012).